\documentclass[12pt]{article}
\usepackage{a4wide}
\begin{document}
\def\d{{\rm d}}
\begin{titlepage}
\begin{flushright}
{\small preprint LPSC-05-114}\\
{\small hep-ph/0511247}
\end{flushright}
\vskip 4cm
%
%
\begin{center}
{\LARGE \bf Double charm hadrons revisited%
\footnote{Based on an invited talk by Fl.~Stancu at the
Mini-Workop ``Exciting Hadrons'', Bled
(Slovenia), July 11-18, 2005}}\\[2cm]
{\large J.-M.\ Richard\footnote{{\it e-mail}:
Jean-Marc.Richard@lpsc.in2p3.fr }}\\
{\small Laboratoire de Physique Subatomique et Cosmologie,}\\
{\small Universit\'e Joseph Fourier--IN2P3-CNRS,}\\
{\small 53, avenue des Martyrs, 38026 Grenoble cedex, France}\\
\vspace{0.5cm}
{\large Fl.\ Stancu\footnote{{\it e-mail}: fstancu@ulg.ac.be}}\\
{\small Institute of Physics, B.5, University of Liege,}\\
{\small Sart Tilman, 4000 Liege 1,Belgium}%
\end{center}

\vskip 1cm

\begin{abstract}\noindent
The dynamics of two heavy quarks inside the same hadron is probed against 
the double charm baryon results of SELEX. This can be seen as a part 
of the mechanism to bind tetraquarks with two heavy quarks and two light 
antiquarks. In the framework of potential models, it is possible to test 
the role of different effects: relativistic corrections, confinement, 
hyperfine forces, etc. It is conjectured that an additional  interaction
rescaled from the nucleon--nucleon system and acting between light quarks
only, can help in bringing extra, possibly required, binding in tetraquarks.
\end{abstract}
\end{titlepage}

\section{Introduction}
The observation of double charmed baryons, $(ccq)$,  by the SELEX
collaboration \cite{Mattson:2002vu,Ocherashvili:2004hi}
has brought new incentive to study hadrons containing
heavy  quarks. The discovery by Belle and BaBar of the $\mathrm{D}_{s,J}$  
states \cite{Aubert:2003fg}, 
of  the anomalously narrow $X(3872)$ meson \cite{Choi:2003ue},
and of the other hidden-charm states $X(3940)$ \cite{Abe:2004zs} 
and $Y(4260)$ \cite{Aubert:2005rm} also stimulated much activity in this field. Some of these states might be interpreted as meson--meson molecules or as multiquark states. 
 
Regarding double-charm baryons, several uncertainties are left in the SELEX 
results: only the 
$(ccd)^+(3520)$ is confirmed, as being seen in two different weak-decay  modes. 
Unfortunately, other peaks, which are candidates for isospin partner, spin or orbital excitation, are far from being established \cite{Engelfried:2005kd}.  There is hope that the final analysis of SELEX data and future experiments could clarify the situation. In particular, the double charm production seen in B-factories to detect charmonium states recoiling against other charmonium states, could possibly lead to final states with double charm hadrons recoiling against  double anticharm systems.

The purpose of the present study is to
investigate whether or not charmonium
$(c\bar{c})$, charmed mesons $(c\bar{q})$ and
baryons $(cqq)$ can be described in a simple
unified picture and lead to predictions for
double-charm baryons $(ccq)$ and for hidden-charm
$(cq\bar{c}\bar{q})$ and double-charm
$(cc\bar{q}\bar{q})$ tetraquarks.

We believe that mixing heavy and light quarks or antiquarks in the same 
system  sometimes favours multiquark binding below the threshold for 
spontaneous dissociation into simpler hadrons. Some attractive effects 
between two heavy quarks, or between two light quarks, might lower the mass 
of a multiquark system without acting on the threshold energy. 

The structure of this paper is a follows. In the next section, we present 
the main dynamical ingredients that play a role in multiquark binding.  
In Secs.\ \ref{se:phe} and  \ref{se:dbc}, we briefly review some of the 
theoretical approaches to double charm baryons.
Section ~\ref{se:tetra} is devoted to an analysis of the latest results 
on open charm tetraquarks.  The last section raises several questions on  
the role of various dynamical effects and, in particular, 
of the light quark dynamics in the binding of tetraquarks.
\section{Aspects of quark dynamics}\label{se:qd}
\subsection{The heavy--heavy effect}
Flavour independence is one of the appealing
features of quark models, directly linked to QCD.
Potentials have been designed, for instance, to
describe simultaneously the $(c\bar{c})$ and
$(b\bar{b})$ spectra. In a flavour-independent
potential used in the context of the Schr\"odinger,
or an improved equation, with
relativistic kinematics, one automatically gets a
\emph{heavy--heavy} effect: a subsystem with
large reduced mass takes better benefit  of the
attraction. This is illustrated by the following
inequalities \cite{Nussinov:1999sx}.
\begin{eqnarray}
(Q\overline{Q})+ (q\bar{q})&\le 2 (Q\bar{q})~, \label{ineg1}\\
(QQq)+ (qqq)&\le 2(Qqq)~,\label{ineg2}\\
(QQ\bar{q}\bar{q})&\le  2(Q\bar{q})\label{ineg3}~.
\end{eqnarray}
While (\ref{ineg1}) and (\ref{ineg2}) are valid
for any value of the heavy-to-light mass ratio
$x=M/m$, (\ref{ineg3}) requires a minimal value
of $x$ for a bound state to occur. In atomic
physics, a sort of flavour independence is also
present, inasmuch as the same Coulomb potential acts on
light and heavy charges. There the inequality (\ref{ineg1})  means
that a system of a protonium and a positronium is
lighter than the hydrogen plus the antihydrogen. The
inequality  (\ref{ineg3}) is also observed: while
the positronium molecule is marginally bound, the
hydrogen molecule lies well below the threshold
of dissociation into two hydrogen atoms.
\subsection{The light--light effect}
In quark physics, one also encounters a 
\emph{light--light} effect, which is not
explicitly included in simple potential models
and thus should be added by hand. Two hadrons,
containing at least one light quark each, have a
long-range strong interaction which is sometimes
attractive and might contribute to binding. The
best known example is the deuteron. The
$(\mathrm{D}\overline{\mathrm{D}}{}^*
+\mathrm{c.c.})$ is another possibility, which
has perhaps been seen in the Belle data \cite{Choi:2003ue}. This
light--light effect also contributes to the inner
dynamics of hadrons with two or more light
quarks. 

In Sec.\ \ref{se:tetra}, 
we shall speculate about the role of an additional meson exchange interaction at
the quark level, as in Ref. \cite{Froemel:2004ea,Julia-Diaz:2004rf}. Only the residual interaction
between light-flavour quarks is admitted to be significant. Accordingly,
such an interaction, rescaled from the nucleon--nucleon interaction can
increase the binding the two interacting mesons.
\section{Phenomenological models}\label{se:phe}
Several attempts have been made to build  potentials that describe 
simultaneously meson and
baryon masses, with a suitable ansatz for going from the quark--antiquark  
to the quark--quark case \cite{Stanley:1980fe}. Some models 
were even extended from the heavy to the light quark 
sector \cite{Bhaduri:1981pn} and can be applied to systems such as charmed 
mesons and double charm baryons and tetraquarks.

For the sake of the discussion, we consider explicitly the potential of  
Bhaduri et al.\ \cite{Bhaduri:1981pn} and the AL1 
potential \cite{Silvestre-Brac:1996bg}, the parameters of which include 
constituent-quark masses, the string tension of a linear confinement, the strength of the Coulomb interaction, and the strength and size parameters of the hyperfine interaction which is a smeared contact term. The Hamiltonian reads
\begin{eqnarray}\label{ham}
&&H= \sum_i m_i
  + \sum_i \frac{\vec{p}_{i}^{2}}{2m_i}
  - \frac {(\sum_i \vec{p}_{i})^2}{2\sum_i m_i}
  + \sum_{i<j} \left[V_{\ell}(r_{ij})+ V_{c}(r_{ij})+ V_{h}(r_{ij})\right]\, ,\nonumber\\
&&V_{\ell}(r_{ij}) = -\frac{3}{16}~\lambda_{i}^{c}\cdot\lambda_{j}^{c} \, ~ (a r_{ij}- b) \, ,\quad
V_{c}(r_{ij}) =-\frac{3}{16}~\lambda_{i}^{c}\cdot\lambda_{j}^{c} \,
\frac{\kappa}{r_{ij}} \, ,\\
&&  V_{h}(r_{ij}) =-\frac{3}{16}\frac {\kappa}{m_i m_j r^2_0} \frac{\exp(-r_{ij}/r_0)}{r_{ij}}
\,\lambda_{i}^{c}\cdot\lambda_{j}^{c}\,
\vec{\sigma_i} \cdot \vec{\sigma_j}~.\nonumber
\end{eqnarray}
In the case of AL1, the smearing parameter $r_0$ depends on the reduced mass of the quark pair.
\section{Double charm baryons}\label{se:dbc}
An extensive  study of the double-charm baryons has been performed  by
Fleck and Richard \cite{Fleck:1989mb}
and a review of the situation at that time can be found in Refs.\ \cite{Savage:1990pr,Fleck:1990ma}. This has been followed by other potential model studies, as
for example, Ref.\ \cite{Silvestre-Brac:1996bg}. These studies suggest that the ground
state
$\Xi^{++}_{cc}(ccu)$
and $\Xi^{+}_{cc}(ccd)$ have a mass around 3.6 GeV
(for more examples, see Ref.\ \cite{Kiselev:2002iy}). In a more recent work
\cite{Vijande:2004at}
an effective scale-dependent strong-coupling constant which distinguishes
between $qq$, $cq$ and $cc$ pairs has been used  to calculate the
spectrum of double charmed baryons, but the ground state
of  $\Xi^{+}_{cc}(ccd)$ was fitted to
the SELEX data \cite{Mattson:2002vu,Ocherashvili:2004hi} at 3520 MeV.
Double-charm baryons have also been investigated in lattice QCD.
Predictions for masses and spin splittings were made in lattice nonrelativistic 
quenched QCD calculations \cite{Lewis:2001iz,Mathur:2002ce} prior to the SELEX experiment.
Recently, quenched lattice calculations
with exact chiral symmetry \cite{Chiu:2005zc} showed an agreement with
the published SELEX data \cite{Mattson:2002vu,Ocherashvili:2004hi}.
There are also studies based on an effective field theory Lagrangian approach
adequate for heavy quarks \cite{Brambilla:2005yk}.

\begin{table}[!h]
\caption{\label{BARYON}
The baryon masses (in MeV)  obtained   obtained from the
 Bhaduri et al. \cite{Bhaduri:1981pn} model, without and with
  hyperfine interaction $V_h$, compared to the experimental values, from PDG \cite{Eidelman:2004wy}
  for single  charm  and from SELEX data~\cite{Mattson:2002vu}
  for double charm baryons.}
  \begin{center}
\begin{tabular}{c|c|c|l}
\hline
Content & without  $V_{h}$ & with $V_{h}$ & Exp.   \\ 
\hline
    &  & 2332 $(1/2^+)$ &  $\Lambda_c(2285) $  \\
$cqq$   &    2500             & 2500 $(1/2^+)$ &  $\Sigma_c(2455)  $ \\
  &                 & 2568 $(3/2^+)$ &  $\Sigma_c(2520)  $ \\
  & & &\\
$ ccq $  & 3693 & 3643 $(1/2^+)$ & $ \Xi_{cc}^+$(3520)\\
 &           & 3724 $(3/2^+)$ & \ \ ?\ \  \\
\hline
\end{tabular}\end{center}
\end{table}

In Table \ref{BARYON}, preliminary estimates of  the masses 
from   the Bhaduri et al.\  potential \cite{Bhaduri:1981pn}
are compared with the experimentally known masses.
The results are shown without and with hyperfine interaction. For single charm baryons we use 
PDG data \cite{Eidelman:2004wy} and for the confirmed double charm baryon the 
SELEX data \cite{Mattson:2002vu}.
Note that the spin and parity are quark
model predictions, not determined experimentally so far, even for 
single charm baryons. 
Our estimate is made with the powerful method of Kamimura et al.\ \cite{Hiyama:2003cu}. So far, we have used a small number of terms in the Gaussian expansion. A  better calculation with a larger basis will be presented elsewhere.
  
The mass of  $(ccd)(1/2^+)$ is found around 3.6 GeV, consistent with several 
previous constituent quark model calculations~\cite{Kiselev:2002iy}
and lattice  
results \ \cite{Lewis:2001iz,Mathur:2002ce}. Note that 
the hyperfine splitting of  $(1/2)^+$ and  $(3/2)^+$ states is about
80 MeV, similar to most quark model studies and lattice calculation 
results, but at variance with the 24 MeV splitting
of Ref.\ \cite{Vijande:2004at}, which seems to be anomalously small.
\section{Tetraquarks}\label{se:tetra}

An understanding of baryons with two heavy quarks could
make it possible to better extrapolate towards 
tetraquarks with double heavy flavour 
\cite{Rosina:2003gg,Gelman:2002wf}.
Let us consider the tetraquarks with open charm $cc \bar q \bar q$ 
which have been extensively studied in the framework
of the Bhaduri et al. potential
\cite{Bhaduri:1981pn} or in some of its improved 
versions \cite{Silvestre-Brac:1996bg}.

Table \ref{TETRA} displays the binding energy $\Delta E =M(cc \bar q \bar q)- M_{th}$ of the lowest state having spin $S=1$ and isospin $I=0$
calculated with the Bhaduri et al. potential and with the AL1 potential. The masses  were obtained with  
two different numerical methods. The threshold mass is
$M_{th}=M(D) + M(D^*)$.

\begin{table}[!h]
\caption{\label{TETRA}
The $I=0$, $S=1$ tetraquark binding energy   (in MeV)
$\Delta E = M(cc \bar q \bar q) - M_{th}$, 
where the  threshold mass $M_{th}$ is calculated with the same model.}
\begin{center}\begin{tabular}{c|c|c}
\hline
Potential & Bhaduri et al. & AL1\\
\hline
Silvestre-Brac \& Semay \cite{Silvestre-Brac:1993ss}  & 19   & 11  \\
Janc \& Rosina \cite{Janc:2004qn}    & $-0.6$ & $-2.7$   \\
\hline
\end{tabular}\end{center}
\end{table}

The tetraquark with spin $S=1$ and isospin $I=0$  appears unbound
in  Ref.~\cite{Silvestre-Brac:1993ss}, where the four-body problem 
is solved by an expansion in a harmonic-oscillator basis up to $N=8$ quanta. 
However, it is bound in the calculations by  Janc and Rosina \cite{Janc:2004qn},
who used a multi-channel variational basis, as presented in 
Ref. \cite{Brink:1998as}, by
including, in particular, meson--meson type of asymptotic channels.  
It was already noted in Ref.\ \cite{Zouzou:1986qh} that the 
meson--meson configurations are important when the stability limit 
is approached. On the other hand, using the mass of 3520 MeV for 
$ccu/ccd$ from SELEX
the mass of the tetraquark $cc \bar u \bar d$ 
becomes about 3905 MeV i. e. some 35 MeV above the 
$D + D^*$ threshold \cite{Gelman:2002wf}, consistent with earlier 
estimates \cite{Rosina:2003gg}.

The picture of a ``two-meson" state in tetraquarks also emerges from
recent SU(3) lattice QCD calculations \cite{Okiharu:2004ve}.  There it is seen
as a flux-tube recombination, known as ``flip-flop",
when a quark and antiquark are near each other. The extension to
pentaquarks has been considered in a more recent calculation
\cite{Okiharu:2005eg}.

There are several ways to increase the binding, all related
to long-range forces.  One concerns the confinement potential. 
The common assumption is that the confinement is two-body, 
although this is too a simplified picture in the light of lattice
calculations \cite{Okiharu:2004ve}. It has been suggested that a three-body
confinement interaction can be introduced as a colour operator 
via the cubic invariant of SU(3) \cite{Dmitrasinovic:2001nu}.
This is a pure algebraic approach, without an underlying physical picture, so far. 
It  can produce an increase of the binding, depending on the sign and strength of the three-body interaction~\cite{Janc:2004qn}. 
\section{Perspectives}\label{se:per}
It is remarkable that stable tetraquarks are predicted from Hamiltonians 
adjusted to 2- and 3-body systems and applied to 4-body systems. 
If the  $(cc\bar{q}\bar{q})$ state exits, it will be accessible to ongoing 
and future experiments.

The prediction deserves further investigation and the natural question is 
whether or not the stability survives changes in the basic assumptions. 
This is the aim of our present and future study.  Several questions can be 
raised, for instance: 
\begin{itemize}
\item
are the relativistic effects important, are they more important in tetraquarks than in mesons and baryons?
\item
is the interaction between quarks pairwise?
\item
in the case of a two-body interaction is the $\lambda_{i}^{c}\cdot\lambda_{j}^{c}$ operator appropriate to describe the colour dependence?
\item
is the linear parametrisation of the confinement adequate?
\item
do we need to introduce asymptotic-freedom type of correction to the strength $\kappa$ of the Coulomb term?
\item
is the chromomagnetic  interaction $V_h\propto \,\lambda_{i}^{c}\cdot\lambda_{j}^{c}\,
\vec{\sigma_i} \cdot \vec{\sigma_j}$ realistic enough to describe hyperfine effects?
\item
do results on stability depend strongly on the assumed regularisation of the hyperfine interaction?
\item
does a tensor type interaction increase the binding in tetraquarks?
\end{itemize}

In addition to these questions,  it seems crucial to us to investigate the role of the light--light effect mentioned in the introduction. This is required by chiral dynamics as well as by empirical evidence.
An observation is that in simple quark models, it is difficult to accommodate simultaneously light and heavy hadrons. Also a long-range hadron--hadron exists, if both hadrons contain light quarks. This suggests to introduce a residual interaction of meson-exchange type, admitted to be significant only  between light quarks. This pedestrian way of implementing chiral symmetry at the quark level leads to interesting predictions. 
According to Refs.~\cite{Froemel:2004ea,Julia-Diaz:2004rf}, the contribution of light pairs represents a fraction of the nucleon-nucleon interaction, so it can be obtained by rescaling 
the nucleon-nucleon interaction to the corresponding hadron--hadron system. In particular,
the $\Xi_{cc} - \Xi_{cc}$ interaction includes a pion exchange component.
Although its strength is only a fraction of the nucleon--nucleon interaction,
\footnote{For example, the light quark fraction of the central part 
and of the spin-isospin part of the long range interaction are 1/9 and 
1/25 respectively \cite{Julia-Diaz:2004rf}.}                       
 a deuteron-like bound state is likely 
to exist between double charm baryons because the kinetic energy has 
a less repulsive effect in $\Xi_{cc} - \Xi_{cc}$ than in $NN$,
 as a consequence of the large mass of  $\Xi_{cc}$. 

By analogy, we expect that in a  $(cc \bar q \bar q)$ system the 
$ \bar q \bar q$ pair should bring an extra attraction
like in  $\Xi_{cc}$ - $\Xi_{cc}$, and possibly
lead to stable tetraquarks against strong decays.

In the future we plan to estimate the contribution of 
the light pairs of quarks to the mass of
$c c \bar q \bar q$ from a realistic nucleon-nucleon interaction, as 
for example the Paris potential \cite{Lacombe:1980dr}.

The $(cc \bar q \bar q)$ state has the unique feature to combine the 
heavy--heavy and light--light effects which are absent in its dissociation
products, and hence offers the best possibility for multiquark binding.


\end{document}